\documentclass[conference]{IEEEtran}
\IEEEoverridecommandlockouts
\usepackage{cite}
\usepackage{amsmath,amssymb,amsfonts}
\usepackage{algorithmic}
\usepackage{graphicx}
\usepackage{textcomp}
\usepackage{xcolor}
\usepackage{pifont}
\usepackage{makecell} 
\usepackage{booktabs}
\usepackage[ruled,vlined,linesnumbered]{algorithm2e}

\def\BibTeX{{\rm B\kern-.05em{\sc i\kern-.025em b}\kern-.08em
    T\kern-.1667em\lower.7ex\hbox{E}\kern-.125emX}}
\begin{document}


\title{MicroIntent: Intent-Based Placement Strategy for Microservice Application in the Compute Continuum Using LLMs}

\author{\IEEEauthorblockN{Koushikur Islam\textsuperscript{1}, Guilherme da Cunha Rodrigues\textsuperscript{1,2}, Bahman Javadi\textsuperscript{1}, Rodrigo N. Calheiros\textsuperscript{1}}
\IEEEauthorblockA{\textsuperscript{1}School of Computer, Data and Mathematical Sciences, Western Sydney University, Australia \\
\textsuperscript{2}Federal Institute of Education, Science and Technology Sul-Riograndense (IFSUL), Brazil \\
k.shohag@westernsydney.edu.au, guilhermerodrigues@ifsul.edu.br, \{b.javadi, r.calheiros\}@westernsydney.edu.au}
}

\maketitle

\begin{abstract}
The placement of microservices in the compute continuum plays a vital role in delivering services that comply with customers' needs, such as reduced latency, storage requirements, quality of service and availability. To achieve customers’ needs in the geographically dispersed architecture of the compute continuum, Service Level Objectives (SLOs) have been largely used in decision-making to place microservices. However, because low-level SLOs increase the barrier to entry for continuum users, placement decisions based on high-level business vocabulary are required if the compute continuum is to be adopted at scale. This paper proposes an architecture for microservices placement decisions in the computing continuum utilizing high-level user intents described in natural language as input. The approach utilizes Generative Artificial Intelligence to translate the intents to low-level SLOs, which are used along with the infrastructure description to decide where different microservices that compose an application must be deployed so that SLOs are met. We implement and evaluate a prototype of the architecture to demonstrate the approach's feasibility.
\end{abstract}


\section{Introduction}
The rise in demand for scalable and distributed software applications has introduced complexity in resource management due to the heterogeneous architecture of modern systems. To address this, software systems have adopted decoupled architectural patterns, enabling deployment across diverse locations. This adaptation, while enhancing scalability, has resulted in more complex management to effectively meet user goals.

Software applications have fueled a paradigm shift toward \emph{microservices}, enabling enhanced scalability, flexibility, resilience, and accelerated delivery of applications. Microservices can be considered a method to decouple software into individual components, providing several advantages, including dynamic scaling, faster roll out, and low latency \cite{sebrecht}.


Modern applications built on microservice architecture use cloud systems for higher scalability, lower cost, and improved availability, but are impacted by network latency due to the limited data centers of the cloud~\cite{odun}. Edge~~\cite{mansouri2021review} computing reduces latency by processing data near its sources in real-time, but the constraints of resources limit scalability, computational capacity, and availability compared to the cloud, which are vital for meeting diverse microservice requirements.

In this context, the \emph{compute continuum}, which integrates edge and cloud computing to address the demands of microservices, has emerged as a novel paradigm to fulfill microservices requirements offering reduced latency, higher throughput along with increased scalability and cost-efficiency~\cite{moreshi}.

The placement of microservices in the compute continuum plays a vital role in the delivery of services that comply with customers' needs, such as reduced latency, storage, Quality of Service (QoS) and availability \cite{samani}. 

To achieve customers' needs in a geographically dispersed architecture of the compute continuum, Service Level Objectives (SLOs) have been largely used in decision-making to place microservices. However, because low-level SLOs increase the barrier to entry for continuum users.

Intent Based Networking (IBN) utilizes user intents to manage network operations \cite{gharbaoui2023intent} which are high-level and abstracted description of a network service along with attributes~\cite{clemm2022intent}. This approach of user defined intents using human comprehensive business vocabulary are required if the compute continuum is to be adopted at scale. Complex attributes for describing user intents often hinder users with limited lower level system knowledge to utilize the scalable resources of compute continuum.



To address these issues, this paper introduces \emph{MicroIntent}, an architecture that provides recommendations regarding microservices placement in the compute continuum based on high-level user intents supplied in natural language. The proposed architecture primarily consists of two layers, including a high-level intent parser and a placement generator. Initially, the high-level intent provided to the parser is translated into actionable SLOs, which are used along with the infrastructure description to recommend the placement in the compute continuum using rule-based matchmaking in the placement generator component.



The key contributions of this paper include:

\begin{itemize}
    \item We propose \emph{MicroIntent}, an architecture that provides placement strategy for microservice application in the compute continuum.
    \item We propose Large Language Model (LLM) based extraction of actionable SLOs from natural language user intents leveraging the Natural Language Processing (NLP) capabilities.
    \item We present experiments conducted on a Minimum Viable Prototype (MVP) demonstrating the feasibility of the approach for placement strategy generation.
    \item Our proposed architecture simplifies microservice placement for users of the complex compute continuum allowing, lowering the entry barrier.
\end{itemize}

The remainder of this paper is organized as follows. Section II discusses relevant papers in regard to microservices placement in the compute continuum. Section III introduces the proposed architecture. Section IV presents the evaluation. Section V concludes the paper and suggests future works.

\section{Related Work}
\label{sec:relatedwork}

Microservice applications deployed across computing continuum are distributed in nature \cite{zafeiro} and the heterogeneous and geographically dispersed characteristics of the compute continuum make the seamless deployment of the applications challenging \cite{balouek}. Existing work primarily focuses on resource management aspect overlooking the entry barrier for user with limited lower-level system knowledge by enabling them to express SLOs as intents.

\subsection{Traditional Intent Acquisition Approaches}
Traditional methodologies for intent acquisition emphasize the ingestion of user intents, typically expressed as lower-level SLOs, to facilitate resource allocation accordingly, posing barrier for users lacking expertise of lower-level configurations.

A fog mesh proposed by Sebrecht et al.~\cite{sebrecht} accepts user intents and parses it into relevant workflows, taking into account the locality of end users and the constraint in the workflow to ensure QoS.  This study does not accommodate natural human comprehensive language as a form of intent establishing an entry barrier for users with limited expertise.

Spillner et al.~\cite{spillner} explore controlling microservices in the compute continuum based on high-level meaningful human and business vocabulary. Despite considering high-level meaningful human vocabulary, the proposed solution manages application performance by adjusting the resources available for services in the nodes where they are placed, rather than suggesting where these services should be placed. 


Morichetta et al.~\cite{schahram} explore load balancing, cost-efficiency, and function coordination in serverless computing, utilizing user intents from stakeholders within the compute continuum. While it notes the use of LLMs for natural language-to-configuration translation, popular in IBN, it does not support natural human language, accepting SLOs directly from users instead.

Filinis et al. \cite{filinis2024intent} present an intent-driven orchestration framework for serverless applications in the computing continuum, allowing users to input intent descriptions and function metadata with keywords (constraints, targets, objectives, properties). However, this method, reliant on lower-level configurations with metadata, limits less-experienced users due to the lack of natural language support.

Jacobs et al. \cite{jacobs} introduced "LUMI," a novel system employing a chatbot interface to translate user-provided natural language intents into low-level configurations for IBN deployment. This approach facilitates automated configuration for operators, reducing expertise barriers. However, the Bi-LSTM-NER framework limits its capacity to process complex, unstructured intents into heterogeneous computing continuum.

\subsection{Service Placement Methods}
To optimize microservice placement across the compute continuum, placement algorithms are widely used which mostly focus on the compute and network resources~\cite{zafeiro}.

Samani et al.~\cite{samani} proposed a method that distributes coordination between services to prevent QoS violation through Service Level Agreements (SLAs). This work also takes into consideration the current capabilities and past credibility of devices in deciding the placement strategy. However, the Proactive Application Placement does not accept user intention to decide on placement, instead prioritizing the Service Level Agreements (SLA's) requirement of the devices.

Mota-Cruz et al.~\cite{mota} proposed two distinct approaches to minimize latency and load balance, which they called ``app-based'' and ``service-based'' placement of microservices. While the app-based approach involves placing all the services sequentially in one app, the  service-based approach takes one service at a time and evaluates the impacts of connected loads to analyze placement.
Although this work highlights the findings to justify different placement approaches based on the application's business context using various algorithms, it does not consider high-level intention of users to decide or suggest the placement in the compute continuum.


\subsection{Research Gap in Existing Work}


Diverse factors, including resource adjustment in the compute continuum, devices' past credibility and present capability from SLA, and intention to workflow conversion based on low-level metrics have been studied in the past to decide placement in the compute continuum. Existing solutions have focused on a more formal input of intents, rather than using a human comprehensive vocabulary. 

To address this limitation, in this paper, we propose a novel generic architecture that aims to fulfill this research gap by providing service placement strategy based on the human comprehensive business vocabulary. Furthermore, the architecture facilitates the integration of diverse placement algorithms tailored to user requirements.

\section{Proposed Architecture}

\begin{figure}[!tb]
\centering
\includegraphics[width=0.75\columnwidth]{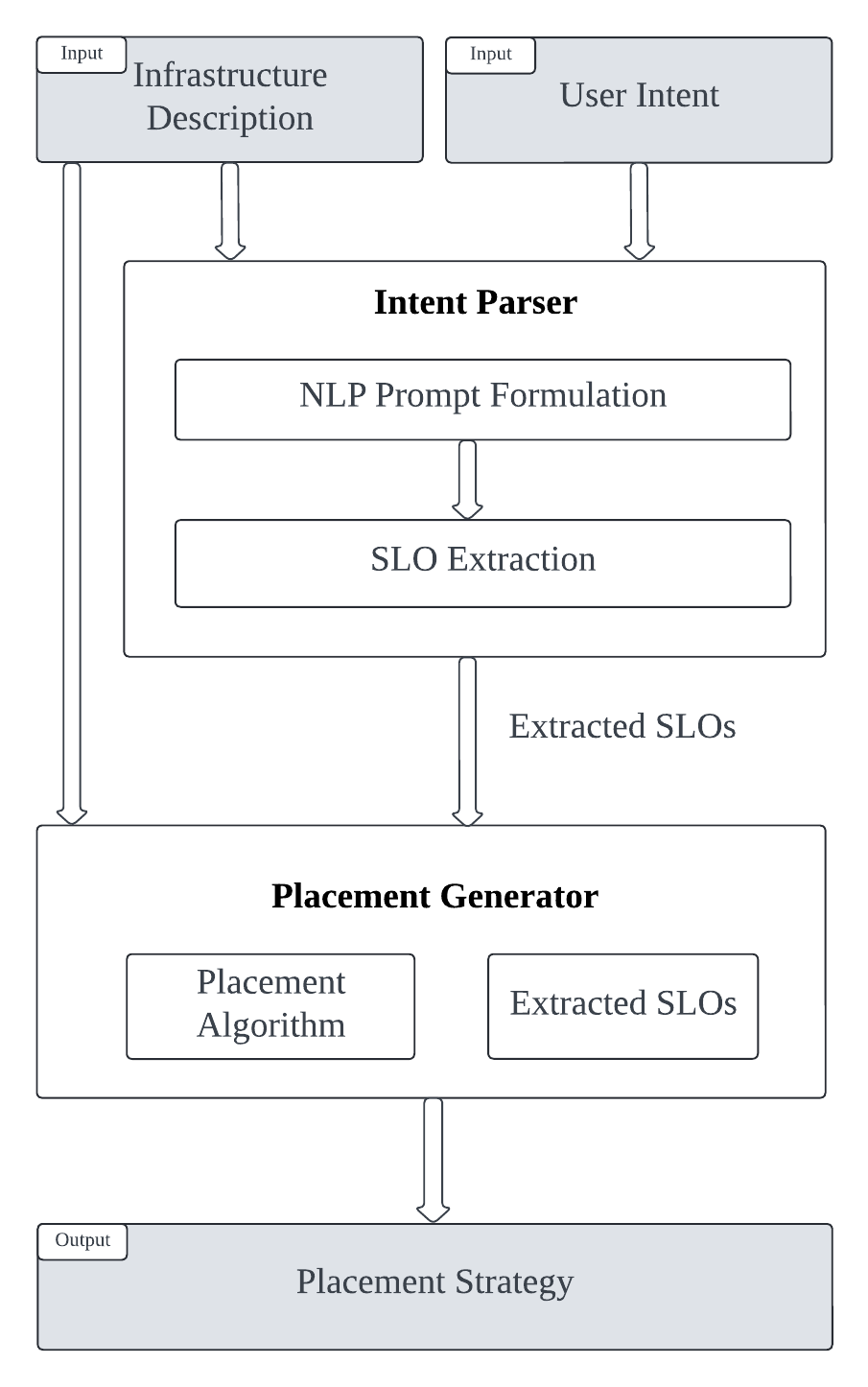}
\caption{The proposed architecture.}
\label{fig:system_architecture}
\end{figure}

Our proposed architecture aims to suggest microservices placement decisions in the compute continuum based on high-level user intents. It is depicted in Figure~\ref{fig:system_architecture} and consists of two main components: \emph{Intent Parser} and \emph{Placement Generator}.

The Intent Parser is responsible for translating intentions, received in natural language, into low-level SLOs for the different components of the services that enable intentions to be realized. The SLOs are recorded in a document in JavaScript Object Notation (JSON) format for consumption by the next module.

The Placement Generator component receives as input both the SLOs generated by the Intent Parser and a document describing the compute continuum infrastructure, also in JSON format. This input is used to generate a recommended placement of the components of microservices applications that meet the desired intent using the placement algorithm.

The expected system input and the modules are discussed in detail in the rest of this section.

\begin{figure*}[!tb]
\centering
\includegraphics[width=0.95\textwidth]{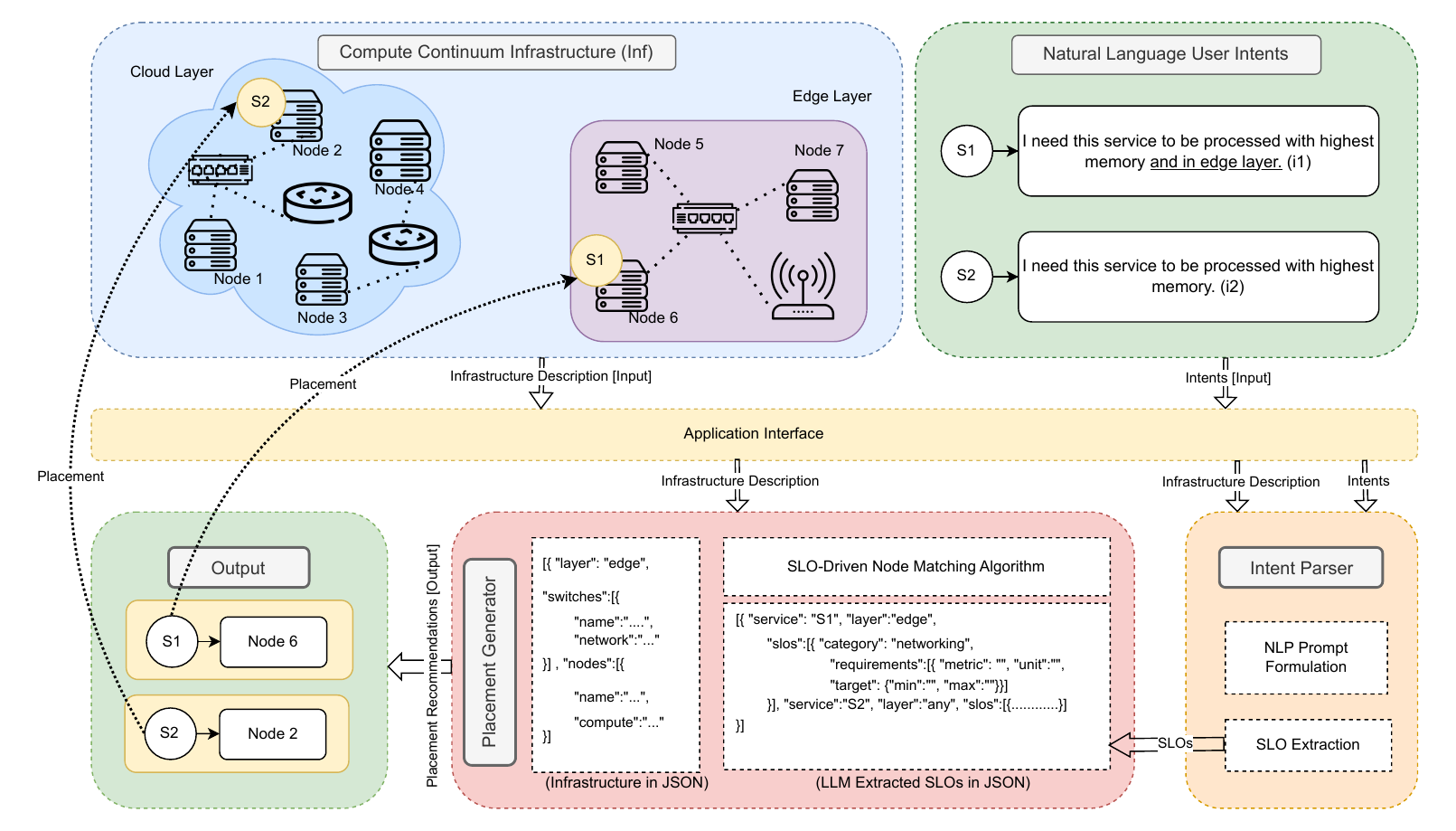}
\caption{Placement recommendation generation process of the proposed architecture.}
\label{fig:system_overview}
\end{figure*}

\subsection{System Inputs}

Two inputs are required by our proposed architecture: a description of the compute continuum infrastructure where the microservices application is to be deployed and the intents regarding the application. 

\subsubsection{Compute Continuum Description}

The compute continuum description is supplied to the system as a JSON document. The document describes each compute node available in the system, as well as the networking details for each compute continuum layer. Computing nodes are described in terms of hardware (i.e. number of CPU cores, GPU, memory available, and storage) and software (Operating System, any specialized software if available). Networking is defined in terms of network bandwidth and latency to other layers in the continuum to which a given layer connects (for example, the latency between the edge and cloud layers). 

%

\subsubsection{User Intents}

User intents are supplied to the system in natural language. The system can interpret indications about the required location (for example, ``service A needs to be deployed in the cloud'') and about resources (for example, ``service B needs high memory''). User intents are translated into either Networking or Compute SLOs from natural language. It is important to notice that the architecture assumes that containers are used to host microservices.

\subsection{Intent Parser}

The Intent Parser translates high-level, natural language intents to SLOs based on the available infrastructure description of the compute continuum. This translation plays a fundamental role in enabling users with limited knowledge of low-level metrics to make informed decisions about where in the compute continuum is suitable to deploy their microservices.

This extraction of SLOs from high-level human comprehensive language is done using Generative Artificial Intelligence (GenAI) tools. The use of GenAI enhances the diversification and versatility of this component as it can extract SLOs considering wider aspects of the user intents and metrics. Besides, it allows the system to match the SLO metrics to the existing infrastructure description. This method relies on LLMs for the translation of intention into actionable SLOs.


The Intent Parser initially collects high-level user intents and augments this description with information about the compute continuum infrastructure. Then, it formulates a prompt to be submitted to an LLM to extract the SLOs matching from user intentions. The prompt formulation process is achieved with the \emph{NLP Prompt Formulation} module, which prepares the prompt alongside the various constraints to ensure the SLO parsing mechanism aligns with user intents.

The constraints help to ensure that no extraneous information disrupts the SLO extraction process by instructing the GenAI to provide the JSON-formatted SLOs following the output structure provided. This is needed because unanticipated intents from users can often lead to metrics being generated in a range rather than in a specific value. 

To address this issue, the prepared prompt instructs the LLM to parse the SLOs in a format that, when necessary, includes two keys (minimum and maximum value for a given metric). In the case of a single value provided, the LLM equals the minimum and maximum range of the metric. Additionally, the constraints provided during NLP prompt formulation ensure that extracted SLOs from high-level user intents match the name and unit with those described in the infrastructure, applying adjustments in the units when necessary.

The Intent Parser implements an \emph{SLO Extraction} module that takes the formulated prompt from the \emph{NLP Prompt Formulation} module to request the GenAI model to provide the intended extracted JSON containing SLOs. It makes the request to and accepts the response from the LLM API as a JSON-formatted container, which is responsible for further refining of data and storing the refined extracted SLOs. The refinement process of the LLM response involves removing redundant information and preparing the data for parsing in a JSON container.

The architecture uses JSON as the data interchange format for extraction purposes. The system output is a JSON document containing extracted SLO information from user intentions, which include information such as service name and target metrics, values, units, and preferred location for deployment (optional).

\subsection{Placement Generator}

The \emph{Placement Generator} module plays a critical role in providing the placement strategy in the compute continuum based on the extracted SLOs from user intents and the infrastructure description. This component receives as input both the document describing the compute continuum infrastructure and the extracted SLOs based on user intents generated by the Intent Parser component.

With this information, Algorithm \ref{alg:placement_algorithm} generates a placement that satisfies the users' intents. Notice that the problem of service placement is an active area of research, and many algorithms were developed for this specific purpose, as discussed in Section \ref{sec:relatedwork}. Because of that, this module enables placement algorithms to be replaced as better ones become available. Furthermore, it allows researchers to test various algorithms using the parsed SLOs to optimize specific parameters they require.

The output of this module is a document mapping each service to the node in the compute continuum, where it is recommended that the service be deployed to meet the user intent. 

With the strategy provided as output, users can use their tools of choice to deploy the services according to the generated recommendations. Alternatively, deployment tools such as Nautilus~\cite{nautilus}, Continuum~\cite{continuum}, and IoTDeploy~\cite{iotdeploy} can be used for this purpose.

Figure~\ref{fig:system_overview} illustrates the placement recommendation process of the proposed architecture. It depicts the compute continuum infrastructure with cloud and edge layers, comprising interconnected compute and networking resources, described in a JSON file. This file, along with natural language intents provided as service-user intent pairs, is supplied into the Intent Parser component.

Figure~\ref{fig:system_overview} also demonstrates a sample JSON structure of SLOs extracted from natural language intents and the infrastructure description. These JSON files are processed by the ``SLO-Driven Node Matching Algorithm" to produce a placement recommendation that meets user objectives. The architecture outputs service-node pairs, directing services to appropriate nodes in the compute continuum for deployment.

\section{Evaluation}

In this section, we discuss the evaluation of our proposed intention-based microservice application placement strategy architecture. Initially, we discuss how an MVP was implemented to enable an evaluation of the architecture in recommending the microservices placement strategy in the compute continuum. Next, we validate our architecture through two scenarios to confirm that the system's decision is based on both the user intents provided as prompts and the characteristics of the compute continuum infrastructure.



\subsection{Implementation}
To enable the proposed architecture to be evaluated, we developed an MVP of the architecture described in Section III. The proposed MVP consists of two primary architectural layers: the Application Interface layer and the Application Programming Interface (API) layer. The Application Interface layer provides an intuitive and user-centric interface, facilitating interaction for users to receive placement strategies. Conversely, the API layer supports the Application Interface by performing the necessary computational processes to generate placement strategies based on user intents and compute continuum description. Both layers are containerized using Docker\footnote{https://docs.docker.com/}, ensuring portability, ease of deployment, and streamlined packaging within a unified solution. The implemented architecture is available on GitHub for reproducibility~\footnote{https://github.com/koushikur-islam/MicroIntent}.


\subsubsection{Application Interface Layer}
\begin{figure}[!tb]
\centering
\includegraphics[width=0.85\columnwidth]{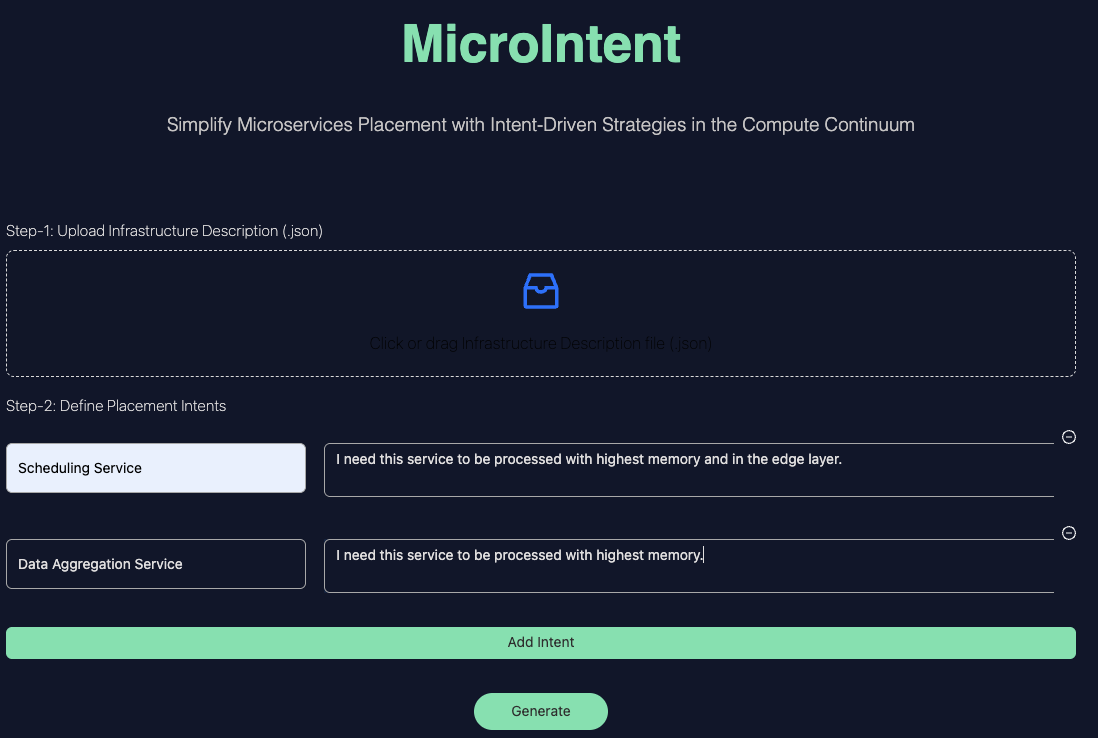}
\caption{Capturing user intents and infrastructure description (.json).}
\label{fig:system_inputs}
\end{figure}

In the proposed MVP, the Application Interface layer is implemented as a web-based user interface, developed using Next.js\footnote{https://nextjs.org/docs}, which provides a user-friendly medium for users to interact with the MVP and obtain placement strategies. Figure~\ref{fig:system_inputs} depicts how natural language intents and compute continuum infrastructure description (.json) are supplied as inputs to the architecture using the user interface. Furthermore, it invokes the necessary APIs from the API layer, delivering the microservices placement strategy to users. 

\subsubsection{Application Programming Interface (API) Layer}
The Application Programming Interface (API) layer is responsible for processing and delivering placement strategies derived from user intents and infrastructure description. This layer is implemented using Flask\footnote{https://flask.palletsprojects.com/en/stable/tutorial/}, a lightweight Python 3 web application framework, to expose an API that provides placement strategies. The API's exposure enables users to integrate it into various platforms of their preference for generating microservices placement strategies. The API layer leverages two core components—namely, the Intent Parser and the Placement Generator—to facilitate the generation of placement strategies.

\subsubsection{Intent Parser} 
In our MVP, this component uses the OpenAI Python library\footnote{https://github.com/openai/openai-python}. In our implementation, no specific LLM is prepared to parse user intents utilizing the NLP capability. Therefore, the LLM provided by OpenAI is used for this setup. The MVP uses the LLM models GPT-4o, GPT-4o-mini and GPT-4  offered by OpenAI. As per the proposed architecture, user intents and server configurations are supplied to this component as two distinct inputs. Both the infrastructure description and the user intent are provided as JSON files consisting of compute and network resources captured in Application Interface layer. This component extracts the required SLOs from the natural language intents utilizing the NLP capabilities of LLM models.



This intent inputs use diverse and natural human language to express intents for application deployment in the continuum. The reason why the Intent Parser uses information from the infrastructure description is to allow the mapping of SLOs and their corresponding units between intents (changing the units if needed) and information about the infrastructure. Therefore, the infrastructure description is also provided for this component.

Once both inputs are provided, the LLM prompt formulation module prepares a message to provide to the  LLM to receive the response with expected SLOs. The prompt supplied to the LLM contains both the intents provided by users and the following description: 

\begin{itemize}
    \item Metrics that are not present in the provided infrastructure description must be excluded.
    \item Metric names from the provided intent must match the ones used in the infrastructure description. 
    \item The `layer' field should default to 'any` unless explicitly mentioned in the intent.
    \item The `slos' field must be a list of two objects, `networking' and `hardware', each containing their respective metrics with the fields `metric', `unit', and `target', where target has two fields, `min' and `max'.
    \item Each service must have a unique `service' field for the service name, `layer' for the intended layer name and the `slos' field, which must correspond to its requirements as defined in the intents.
    \item When intent has a latency metric specified to be reduced between two services, then choose the same layer for those services and put latency equal to zero (0).
    \item Latency is ignored (set as zero) between two nodes in the infrastructure description within the same layer.
    \item The output JSON file should have the following fields:

    \begin{itemize}
        \item `service': Places the name of intended service.
        \item `layer': Name of the intended layer.
        \item `slos': Must be separated into two distinct categories:
        \begin{itemize}
            \item `networking': consists of an array which includes metrics like `latency', `bandwidth', or any other metrics related to network performance.
            \item `hardware': consists of an array that includes metrics such as `memory', `cpu-cores', `gpu-cores', `operating-system', or any other metrics related to hardware or system properties.
        \end{itemize}
    \end{itemize}
    
    \item Both networking and hardware description consist of the following fields:

    \begin{itemize}
        \item `metric': Name of the metric from the intent, such as `latency' for `networking' and `memory' for `hardware'.
        \item `unit': Name of the unit for the intended metric. The `unit' field must follow the rules below. If no unit is specified in the intent, then the unit from the matching metric from the infrastructure description provided is assumed. If a unit is specified in the intent, then the unit is matched to the corresponding unit of the metric from the infrastructure description provided. If required, it is converted to the one specified in the infrastructure description.
         \item `target': The target is the value(s) of the intended metric. This can be both numeric and non-numeric on a high level. The `target' field consists of two values as follows:
    \end{itemize}
    
     \item `min': The minimum value of the intended metric. For numeric ranges, this is the lowest value; for non-numeric descriptors (e.g., low, high, average), it is taken as the minimum matching metric in the specified layer, or in the first layer if none is specified.  
    \item `max': The maximum value of the intended metric. For numeric ranges, this is the highest value; for non-numeric descriptors, it is taken as the maximum matching metric in the specified layer, or in the first layer if none is specified.

    \item `related\_services': This is an array of services that are associated with the requested service for networking purpose. For example, if Service A requires minimum latency with Services 2 and 3 then this array should contain Service 2 and Service 3. It should produce the array where services are connected to another service.
\end{itemize}

The NLP Prompt Formulator is responsible for creating an LLM request from the OpenAI provided GPT-4 model. The response from this request is extracted to a JSON file using Python libraries to trim additional text in the received response. Once the output is received and cleaned, it is saved as another JSON file to serve as input to the Placement Generator.

\subsubsection{Placement Generator}

The implementation of this component in our MVP receives two input JSON files: the JSON file that contains the processed output of the LLM (received from the Intent Parser) and the infrastructure description file, which is received as user input. As a placement algorithm for our MVP, we implemented algorithm listed in Algortihm~\ref{alg:placement_algorithm}, which iterates over each service to be deployed. For each service, it will try to locate a suitable node. The first step regards elimination of layers that are not suitable to the service (Lines 5-9). This is because there might be a requirement for the service to run on a specific layer, in which case the recommendation will be constrained to nodes on such layer. Afterwards, it iterates over each identified SLO to identify nodes that meet them: if there are networking requirements (Lines 12-15), the capacity of switches is checked, and nodes attached to those that meet bandwidth and latency requirements are marked as potential nodes. In the case of hardware requirements (Lines 16-19), the capacity of individual nodes is checked, and all those that meet SLOs are marked as potential nodes. Line 20 filters nodes initially identified, so only those that both belong to the selected layer and meet SLOs remain on the list. The algorithm finishes if no node meets all the requirements of any service (Line 22) or after all the services are mapped (line 25).

Because the unanticipated intentions of users can lead to the generation of SLOs with variable values having both numeric and non-numeric for the matrices, our algorithm addresses both numeric and non-numeric values for metrics when determining if a given node can accommodate a given service. 

Our algorithm also considers both minimum and maximum values for each numeric. In our prototype, the only non-numeric metric explored is the location (i.e., whether the node is in the cloud or in the edge and whether a service needs to be deployed specifically in the cloud or in the edge).

\begin{algorithm}
\caption{SLO-Driven Node Matching Algorithm}
\label{alg:placement_algorithm}
$placement\_suggestions \gets []$\\
\For{$service$ in $Services$}{
    $selected\_layer \gets []$\\
    $selected\_nodes \gets []$\\
    \If{$service.layer$ is undeclared}{
        $selected\_layer.add(all\_layers)$\\
    }
    \Else {
        $selected\_layer.add(service.layer)$\\
    }
    $selected\_nodes.add(selected\_layer.nodes)$\\
    \For{slo in service.slo}{
        $potential\_nodes \gets []$\\
        \If{$slo.category$ is ``networking''}{
            \For{$switch$ in $selected\_layer.switches$}{
                \If{slo is met by switch}{
                     $potential\_nodes.add(switch.nodes)$\\
                }
            }
        }
        \If{$slo.category$ is ``hardware''}{
            \For{$node$ in $selected\_nodes$}{
                \If{slo is met by node}{
                     $potential\_nodes.add(node)$\\
                }
            }
        }
       $selected\_nodes \gets (selected\_nodes \cap potential\_nodes)$\\
    }
    \If{$selected\_nodes$ is empty} {
        return fail\\
    }
    \Else{
        $placement\_suggestions.add$ $(service, selected\_nodes[0])$\\
    }
}
Return($placement\_suggestions$)
\end{algorithm}

Once the mapping of services to nodes is completed, an output file is generated, also as a JSON file. The file contains an array of mapping objects. Each of these objects contains two key-value pairs: a key `service', whose value is a string containing the service name, and a key `node' whose value is a string containing the identification of a node. An example of such output is depicted in Figure~\ref{fig:experimental_validation} consumed by Application Interface layer.


\subsection{Validation} 
This section describes experiments we conducted to validate the output generated by our MVP, which in turn enables us to conclude that the MVP successfully generated a mapping of services to resources based on both the user intents provided as prompts and the characteristics of the compute continuum infrastructure. 

The proposed architecture takes both service-tagged intentions and server node configurations in the Application Interface layer. Placement suggestions received from the API layer directly map the nodes, along with the intended service to be deployed on each node. Therefore, to validate the placement suggestion with an intent and its accuracy, both the node configuration and the output placement JSON files are required. Furthermore, it is important to note that the architecture considers only single-point (bandwidth, queue depth, port-utilization, congestion-level) network metrics, as services placed within the same layer under the intent of minimizing network requirements are assumed to have zero point-to-point (latency, round-trip time, jitter) metrics.

To validate the deployment strategies generated by the proposed architecture, experiments were conducted across two distinct scenarios. The first scenario focuses on changes in decision-making based on changes in user intent. In contrast, the second scenario focuses on changes in the decision-making based on changes in the compute continuum descriptions.

\subsubsection{Scenario 1: Compute Location Verification}

In this scenario, we demonstrate the system's reaction to changes in intent in terms of placement requirements. The compute continuum in Scenario 1 is described by the infrastructure description \( \text{Inf}_1 \). The application consists of two microservices, represented as the set \( S = \{S_1, S_2\} \), where \( S_1 \) denotes the Scheduling Service and \( S_2 \) represents the Data Aggregation Service. 

User intents, defined as \( I = \{i_1, i_2\} \), are provided to the MVP as depicted in Figure~\ref{fig:system_inputs}:

\begin{itemize}
    \item \(i_1\) = I need this service to be processed with the highest memory and in the edge layer.

    \item \(i_2\) = I need this service to be processed with the highest memory.
\end{itemize}

For the user intents, \( i_1 \) specifies edge-layer deployment and high memory capacity for \( S_1 \), and \( i_2 \) requests the highest memory capacity across the continuum for \( S_2 \).

\begin{table}[!t]
\centering
\caption{Compute Continuum Infrastructure Description (\( \text{Inf} \))}
\begin{tabular}{@{}lccccc@{}}
\toprule
Node & \makecell[t]{vCPU \\ (Cores)} & \makecell[t]{vGPU \\ (Cores)} & \makecell[t]{Memory \\ (GB)} & \makecell[t]{Location } & \makecell[t]{Switch} \\ \midrule
\( n_1 \) & 16 & 8 & 64 & cloud & \( w_1 \) \\
\( n_2 \) & 32 & 16 & 128 & cloud & \( w_1 \) \\
\( n_3 \) & 64 & 32 & 256 & cloud & \( w_2 \) \\
\( n_4 \) & 4 & 2 & 16 & edge & \( w_3 \) \\
\( n_5 \) & 8 & 4 & 32 & edge & \( w_3 \) \\
\( n_6 \) & 16 & 8 & 64 & edge & \( w_3 \) \\ \bottomrule
\end{tabular}
\label{tab:infrastructure_description}
\end{table}

The infrastructure description, as presented in Table~\ref{tab:infrastructure_description}, outlines the computational resources distributed across the edge and cloud layers, with each node associated with a specific switch. This description constitutes the comprehensive compute continuum description, denoted as \( \text{Inf}_1 = \text{Inf} \), as detailed in Table~\ref{tab:infrastructure_description}.

The infrastructure \( \text{Inf}_1 \) supplied through a JSON file, encompasses a set of compute nodes \( N = \{n_1, n_2, n_3, n_4, n_5, n_6\} \), partitioned into cloud nodes \( N_{\text{cloud}} = \{n_1, n_2, n_3\} \) and edge nodes \( N_{\text{edge}} = \{n_4, n_5, n_6\} \). The network topology comprises switches \( W = \{w_1, w_2, w_3\} \), with \( w_1, w_2 \in W_{\text{cloud}} \) and \( w_3 \in W_{\text{edge}} \), which collectively facilitate inter-connectivity among the compute nodes.

\begin{figure}[!t]
\centering
\includegraphics[width=\columnwidth]{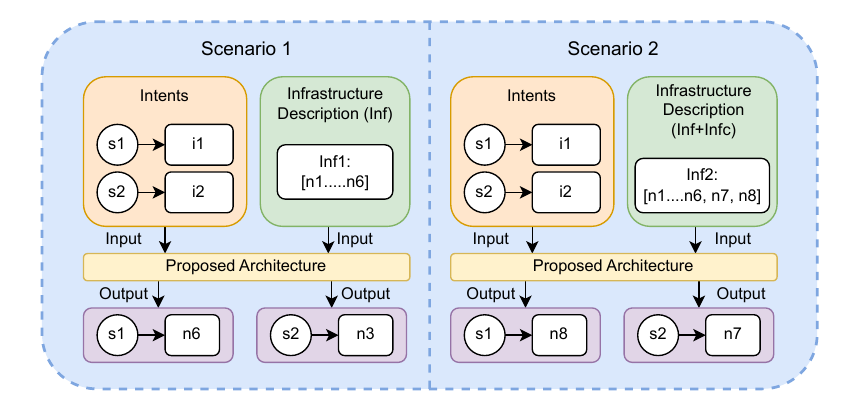}
\caption{Experimental results depicting changes in placement decision driven by user intents and infrastructure description for Scenarios 1 and 2 in the proposed architecture.}
\label{fig:experimental_validation}
\end{figure}

Upon processing the service-specific intents \( I \) and the infrastructure description \( \text{Inf}_1 \), the MVP accurately generates a placement strategy, mapping \( S_1 \to n_6 \in N_{\text{edge}} \) and \( S_2 \to n_3 \in N_{\text{cloud}} \), as illustrated in Figure~\ref{fig:experimental_validation}. This placement satisfies the intents: \( i_1 \) prioritizes edge-layer deployment and high memory capacity (64 GB) for the Scheduling Service (\( S_1 \)), and \( i_2 \) requires the node with the highest memory capacity (256 GB) for the Data Aggregation Service (\( S_2 \)) within the continuum. These results validate the architecture’s capability to align placement decisions with both location-based and resource-based intents within the compute continuum defined by \( \text{Inf}_1 \).

\subsubsection{Scenario 2: Compute Resource Verification}
This scenario validates the architecture's ability to handle compute requirements based on the changes in the compute continuum infrastructure description. The compute continuum, consisting of compute and network resources, is subject to change. We supplied the same intents as in Scenario 1 depicted in Figure~\ref{fig:system_inputs}. However, we added two compute nodes with higher memory capacity to change the infrastructure description to validate the architecture's intent satisfaction approach. Table~\ref{tab:infrastructure_description_2} depicts the addition of two compute nodes with higher memory. Therefore, the infrastructure \( \text{Inf}_2 = \text{Inf} + \text{Infc} \), documented in Tables~\ref{tab:infrastructure_description} and~\ref{tab:infrastructure_description_2} through a JSON file is supplied to the MVP.

\begin{table}[!t]
\centering
\caption{Additional Compute Nodes for Compute Continuum (\( \text{Infc} \))}
\begin{tabular}{@{}lccccc@{}}
\toprule
Node & \makecell[t]{vCPU \\ (Cores)} & \makecell[t]{vGPU \\ (Cores)} & \makecell[t]{Memory \\ (GB)} & \makecell[t]{Location } & \makecell[t]{Switch} \\ \midrule
\( n_7 \) & 128 & 64 & 512 & cloud & \( w_1 \) \\
\( n_8 \) & 32 & 16 & 128 & edge & \( w_3 \) \\ \bottomrule
\end{tabular}
\label{tab:infrastructure_description_2}
\end{table}

This time, the MVP generated different output for the placement than in Scenario 1, although the supplied intents remained unchanged. Nevertheless, as expected, the changes to the compute resources infrastructure description have led the MVP to make a different placement decision. Thus, the MVP could accurately select \(n8\) for \(S1\) with the highest memory in the edge layer and \(n7\) with the highest memory in the continuum for \(S2\) after supplying an updated infrastructure description satisfying the user intents.

These two scenarios, combined, demonstrate that our MVP is capable of (i) reacting to subtle changes in the intent and adjusting the placement accordingly; and (ii) reacting to changes in the infrastructure and adapting the placement accordingly, achieving the goals of the proposed architecture.

\section{Conclusion and Future Work}


To reduce the barrier to the adoption of the continuum, intent-based microservice application placement in the compute continuum plays a significant role. Although previous studies have analyzed intention-based microservice placement in the continuum, the necessity of high-level human language as an intention for the placement is overlooked. This study proposed an architecture that leverages Generic Artificial Intelligence and LLM capabilities to decode high-level user intents to lower-level technical requirements aiming to place services in the continuum. Experiments on an MVP demonstrated  architecture can successfully react to changes of intent and in the infrastructure when making placement decisions.

In the future, we aim to enhance our proposed MicroIntent architecture by incorporating intent conflict management and feedback mechanism to improve the user satisfaction. We also plan to introduce LLM models trained with datasets that focuses on extracting SLOs based on human comprehensive language to enhance the application placement in decision the compute continuum.

\bibliographystyle{ieeetr}
\bibliography{references}

\end{document}